# Quantitative description of complex magnetic nanoparticle arrays

Mohammad Reza Zamani Kouhpanji[1,2] and Bethanie J H Stadler[1,3,*]

[1]Department of Electrical and Computer Engineering, University of Minnesota Twin Cities, Minneapolis, USA; [2]Department of Biomedical Engineering, University of Minnesota Twin Cities, Minneapolis, USA; [3]Department of Chemical Engineering and Materials Science, University of Minnesota Twin Cities, Minneapolis, USA;

**Abstract**

First-order reversal curves (FORC) measurements are broadly used for characterization of complex magnetic nanostructures, but a robust framework for quantitative analysis of the FORC data is still obscure despite numerous studies over decades. In this paper, we first establish a framework for extracting quantitative parameters from the FORC measurements conducted on samples including a single type of the magnetic nanostructures to interpret their magnetic properties. We then generalize our framework for quantitative characterization of the samples including multiple types of the magnetic nanostructures to determine the most reliable and reproducible parameter for detailed analysis of the samples. Our approach provides an insightful path for more accurate quantitative description of complex magnetic nanostructures including various different magnetic subcomponents and/or magnetic phases.

**Introduction**

Advancement of nanotechnology has extensively expedited the emergence of novel magnetic nanostructures, such as magnetic nanowires (MNWs), in various research areas, including medical treatment [1]–[4], environmental science [5], [6], and quantum devices [7]–[10]. The magnetic nanostructures have opened numerous opportunities for scientists in different disciplines such as nanomedicine, molecular biology [11]–[13], applied physics, and nanostructured materials [14]–[19]. A unique benefit of magnetic nanostructures is that they can be excited indirectly by implementing an external field, regardless of their surroundings [20], a critical key in quantum communication and biolabeling applications. In all of these applications, it is crucial to know the characteristics of the magnetic nanostructures, which may inhibit or enhance their use depending on the application. Unfortunately the high yielding nanofabrication processes of the magnetic nanostructures do not allow perfectly identical production leading to variation in their magnetic characteristics and functionalities. Besides that, due to lack of a coherent framework for data extraction and analysis, current techniques for the quantitative characterization of magnetic nanostructures are inefficient both at the research level and at the industrial level.

The magnetic nanostructures have been characterized by measuring their magnetization using the major/minor hysteresis loops and/or first-order reversal curves (FORC). Considering speed and accuracy as limiting criteria, the hysteresis loops measurements provide significantly limited information compared to FORC while it is much faster than FORC. Typically, hysteresis loops provide the saturation magnetization and the coercivity that they are sufficient to describe a magnetic nanostructure that it contains only a single type uncorrelated magnetic subcomponents, such as arrays of non-interaction MNWs. However, they fail to fully describe complex magnetic nanostructures, such as arrays of interacting MNWs. On the other hand, FORC offers a comprehensive insight for qualitative and partially quantitative interpretation of any complex magnetic nanostructure. That is because FORC provides plenty of information only by a single measurement that cannot be achieved using the hysteresis loops measurements. Several works have employed different theoretical models, such Mean-field model, to interpret the information in FORC diagrams [21], [22]. Due to computational limitations, these models consider perfect arrangement of building blocks in the magnetic nanostructures with homogenous properties, which they are not matched with experiments. Therefore, to describe the complex magnetic nanostructure and analyze their functionality, it is essential to step beyond the conventional data analysis and representation of FORC data to precisely describe the magnetic nanostructures.

---

[*] Corresponding author, email: stadler@umn.edu

Experimentally, FORC measurements begin by applying a large magnetic field to ensure the positive saturation of a sample. Next, the applied field, H, is reduced to a predefined field, known as a reversal field, $H_r$, and the moment of the sample is measured while H is retuned to positive saturation, see Figure 1a. This process is repeated with decreasing $H_r$ to collect a family of magnetization curves, $M(H, H_r)$, as a function of reversal field and applied field. The FORC distribution is defined as the second derivative of the magnetization with respect to reversal field and applied field, as follows:

$$\rho = -\frac{1}{2}\frac{\partial^2 M(H, H_r)}{\partial H \partial H_r} \tag{1}$$

In FORC analysis, $\rho$ is plotted as a heat-map with the axes representing the coercive field (x-axis, $H_c = \frac{1}{2}(H-H_r)$) and the interaction field (y-axis, $H_u = \frac{1}{2}(H+H_r)$).

Traditionally, the quantitative analysis of the FORC data is done using projection of FORC distribution on the $H_c$-axis and $H_u$-axis, called coercivity distribution ($P_{Hc}$) and interaction distribution ($P_{Hu}$), respectively, see Figure 1a. Notice, the quantitative analysis means quantifying the coercivity and the interaction fields, not quantifying the amount of the magnetic components [21], [23], [24]. These definitions for quantifying the amount of the magnetic components have a critical issue, indeed they are very misleading. For example, if a magnetic nanostructure contains both interacting and non-interacting magnetic subcomponents, the interaction field distribution of the magnetic nanostructure only represents the interaction field distribution of the interaction subcomponents regardless of the amount of the non-interaction subcomponents. That is because the non-interaction subcomponents have zero interaction field. The same limitation persists for the coercivity distribution, in which it discards the nearly zero coercivity subcomponents. As a result, the amount of the magnetic subcomponents cannot be predicted accurately.

To overcome these shortcomings, one solution is to project the FORC distribution on the $H_r$-axis and H-axis, see Figure 1a. The projection of the FORC distribution on the $H_r$-axis can be calculated by taking an integral from Eq. (1) as follows

$$\begin{aligned} P_{H_r}(H_r) &= \int_{H_r}^{\infty} \rho(H, H_r) dH = -\frac{1}{2}\frac{\partial M(H, H_r)}{\partial H_r}\bigg|_{H=\infty} + \frac{1}{2}\frac{\partial M(H, H_r)}{\partial H_r}\bigg|_{H=H_r} \\ &= 0 + \frac{1}{2}\frac{\partial M(H, H_r)}{\partial H_r}\bigg|_{H=H_r} = \frac{1}{2}ISFD \end{aligned} \tag{2}$$

According to Eq. (2), the projection on the $H_r$-axis determines the difference between the magnetic moment of two sequential reversal curves. Equivalently, it is the residual magnetic moment that it is known to be the irreversible switching field distribution (ISFD). Using the ISFD for quantifying the amount of the magnetic subcomponents have two major advantageous compared to the $P_{Hc}$ and $P_{Hu}$. First, since magnetic switching between two stable magnetic equilibriums is always irreversible, the ISFD parameter contains all magnetic subcomponent responses regardless of their amount. Second, it can be determined by measuring the initial magnetization curves that can significantly accelerate the measurements. An equivalent parameter to the ISFD is the backfield remanence coercivity (BRC), see Figure 1b, which is the residual magnetization at zero applied field. The BRC can be determined by taking a derivative of the magnetization at the zero applied field, which is known to be backfield remanance magnetization (BRM), see Figure 1b. It is essential to emphasize that the only difference between the BRC and the ISFD is that the BRC shows the residual magnetization between two sequential reversal curves at the zero field while the ISFD shows the residual magnetization between two sequential reversal curves at the reversal field. For any magnetic nanostructure with purely linear magnetic response, the ISFD and the BRC are similar. However, due to nonlinearities that mainly induced by the interaction fields, these two parameters are not necessarily similar. Therefore, to comprehend our study, we also consider both BRM and BRC parameters in the following sections. Furthermore, the projection of the FORC distribution on the H-axis can be calculated by taking an integral from Eq. (1) as follows

$$P_H(H) = \int_{-\infty}^{H} \rho(H, H_r) dH_r = -\frac{1}{2}\frac{\partial M(H, H_r)}{\partial H}\bigg|_{H_r=H} + \frac{1}{2}\frac{\partial M(H, H_r)}{\partial H}\bigg|_{H_r=-\infty} \quad (3)$$
$$= \frac{1}{2} RSFD + \frac{1}{2}\frac{\partial M_{lower}(H)}{\partial H}$$

According to Eq. (3), the projection on the H-axis determines the initial slope of the reversal curves (the first term) plus the derivative of the lower branch of the hysteresis loop (the second term). Note, the first term is the spontaneous changes of the magnetization with respect to the field, which is the reversible switching field distribution (RSFD), see Figure 1b.

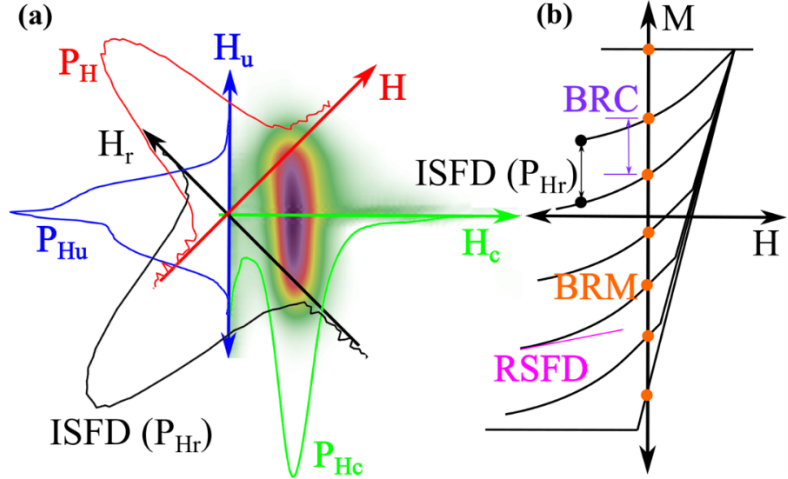

Figure 1: Schematics of FORC distribution (a), and FORC diagram (b). In subfigure (a), the blue curve is the projection on the $H_u$-axis, the red curve is the projection on H-axis, the green curve is the projection on the $H_c$-axis, and the black curve is the projection on the $H_r$-axis.

In the following sections, we extract and scrutinize the aforementioned quantitative parameters ($P_{Hc}$, $P_{Hu}$, $P_{Hr}$, $P_H$, BRM, and BRC) concealed in FORC diagrams via a rigorous statistical analysis. To do so, we practice our approach on FORC diagrams of various types of MNWs arrays and their various combinations. We chose the MNWs arrays because changing their sizes and the interwire distances provide the opportunity to engineer the aforementioned parameters helping to illustrate the nature of those parameters. We then extracted those parameters for the individual types and the combinations. The results are discussed to determine the most reliable and reproducible parameter for quantitative analysis of the volume ratios and types of the MNWs in the combinations.

**Experimental and statistical approaches**

Iron cobalt (FeCo) MNWs were electrodeposited into track-etched polycarbonate templates with a broad range of diameters (fill factors)— 30nm (0.5%), 50nm (1.0%), 100nm (2.0%), and 200nm (12%)— at the room temperature. The electrolyte consists of 0.4M boric acid, 1mM malonic acid, 0.3M ammonium chloride, 0.3mM sodium dodecyl sulfate, 6mM ascorbic acid, 0.2M iron sulfate, and 0.1M cobalt sulfate at pH 3. The concentration ratio of the iron sulfate to cobalt sulfate was chosen 2:1 in order to achieve Fe to Co atomic ratio of 2:1, $Fe_{65}Co_{35}$, which was shown to have the highest saturation magnetization [25]. Figure 2 shows the FORC distributions of the four types of the MNWs measured along the easy axis, parallel to the MNWs axis. Samples with the smallest diameter (fill factor), 30nm (0.5%), had inter-wire distances of ~450nm with inter-wire distance to diameter ratios of 15. This large inter-wire distance presents a fairly symmetric rectangular magnetic hysteresis loop with a localized FORC distribution, which has been predicted by micromagnetic simulations when there are negligible magnetic interactions [26]. By contrast, as the inter-wire distance to diameter ratio decreases, the MNW stray fields interact, leading to a sheared hysteresis loop with a vertically broadened FORC distribution, see Figure 2d. For example, the samples with diameter (fill factor) of 200nm (12%) had inter-wire distances of ~556nm and inter-wire distance to diameter ratios of 2.78, where the resulting FORC distribution broadens vertically indicating a large interaction field between the MNWs. The inter-wire distance to diameter ratio for samples with diameters of 50nm and 100nm is 9 and 7.3, respectively.

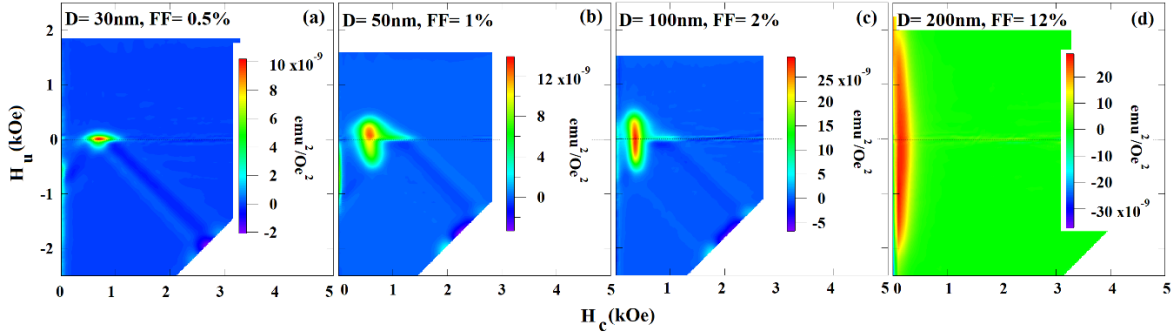

Figure 2: the FORC distributions for different types of the FeCo MNWs with diameter (fill factor) of (a) 30nm (0.5%), (b) 50nm (1%), (c) 100nm (2%), and (d) 200nm (12%).

Several combinations were created with at least two different types of the MNWs, and FORC measurements were repeated, Figure 3. We first extract the $P_{Hc}$, $P_{Hu}$, $P_{Hr}$, $P_H$, BRM, and BRC parameters for the individual samples and combinations. To analysis the reliability and reproducibility of the extracted parameters for quantitative description, the parameters for the combined samples were fit to the corresponding parameters of the individual types of MNWs. The fitting quality was evaluated using the root mean square (RMS) error of the difference between the "experimental data" and "recreated curve", which is the weighted summation of the corresponding parameters of the individual types. The RMS error was minimized to find the optimum weights that they give the volume ratio of the each type of the MNWs in the combination.

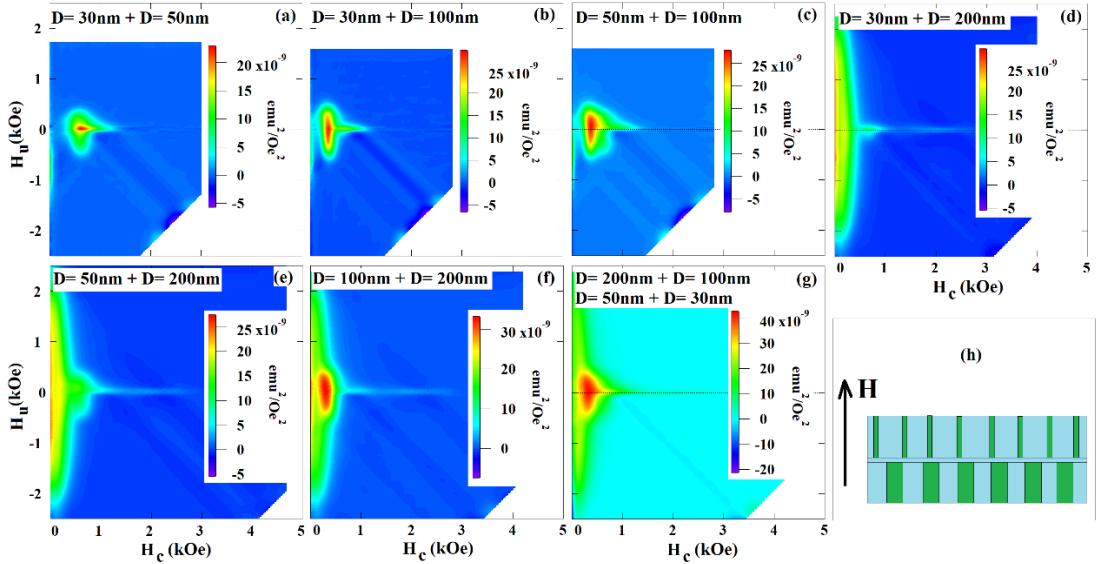

Figure 3: The FORC distribution of combined samples (according to the legends) shows qualitative, not quantitative, differences. Subfigure (h) schematically shows the FORC measurements, where the field direction was parallel to the MNWs (green rectangles) inside the track-etched polycarbonate (blue boxes).

**Results**

Here, we plot the $P_{Hc}$, $P_{Hu}$, $P_{Hr}$, $P_H$, BRM, and BRC parameters for the individual MNWs and their combinations. We meticulously describe the features that these parameters represent to establish a framework for optimizing their efficiency for achieving the best quantitative description of any magnetic nanostructure. The fitting quality and volume ratios will be discussed in the discussion section.

Figures 4 and 5 depict the $P_{Hc}$ and $P_{Hu}$ distributions calculated by taking the integral from the FORC distributions over the $H_u$ and $H_c$ axes, respectively. The location of the peaks in Figure 4 shows the average coercivity of the MNWs, where 30nm (0.5%) MNWs yields the maximum value (~ 0.685 kOe) and 200nm (12%) MNWs have the minimum value (~ 0.163 kOe). These distributions have extra features (e.g. local minima in the top row of Figure 4) compared to the $P_{Hr}$ and/or BRM while they typically do not have any physical interpretations. They are mainly the measurement noises that are exaggerated in the FORC distributions due to data processing, which have been extensively discussed previously [22], [27]–[31].

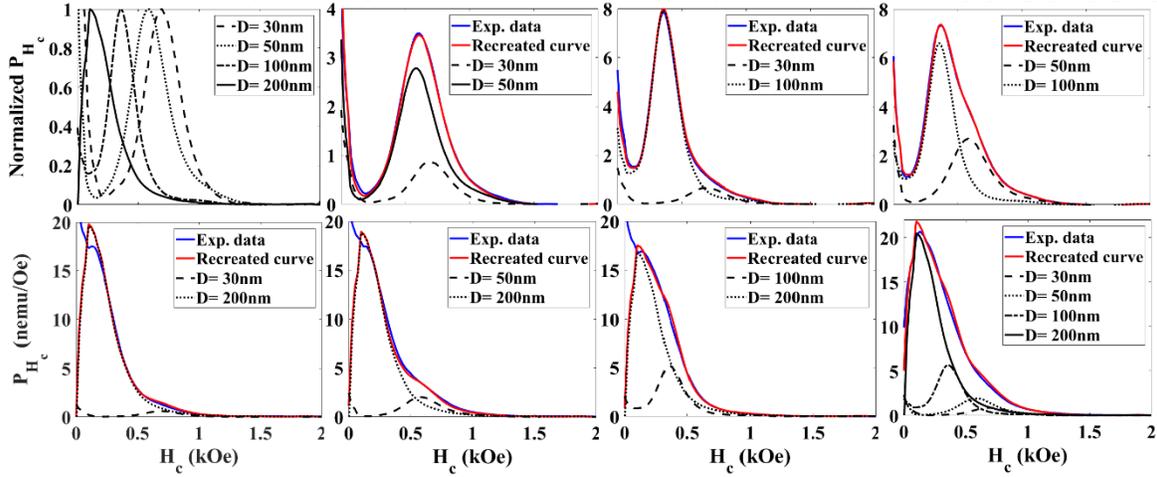

Figure 4: Coercivity distribution ($P_{Hc}$), determined by taking an integral over $H_u$ from FORC distribution, for individual and different combinations of the MNWs as indicated in the legend.

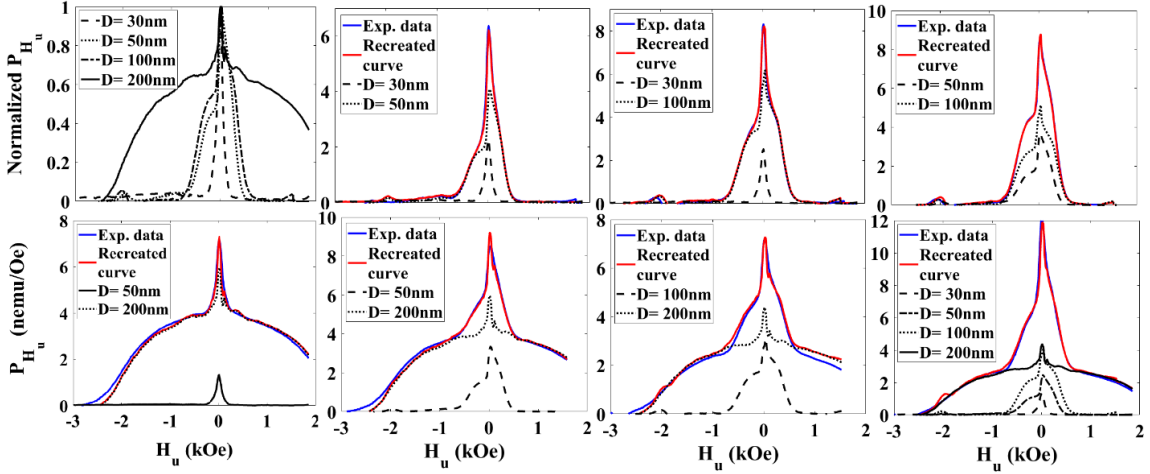

Figure 5: Interaction distribution ($P_{Hu}$), determined by taking integral over $H_c$ from the FORC distributions, for individual and different combinations of the MNWs as indicated in the legend.

The $P_{Hr}$ was calculated by taking an integral of the FORC distribution over all applied fields, Figure 6. The $P_{Hr}$ peak indicates the average coercivity of the sample, similar to $P_{Hc}$ distributions. For combined samples, the ISFD deforms according to the volume ratio and coercivities of the samples in the combination. The shift of the peaks indicate the amount of each MNW relative to another.

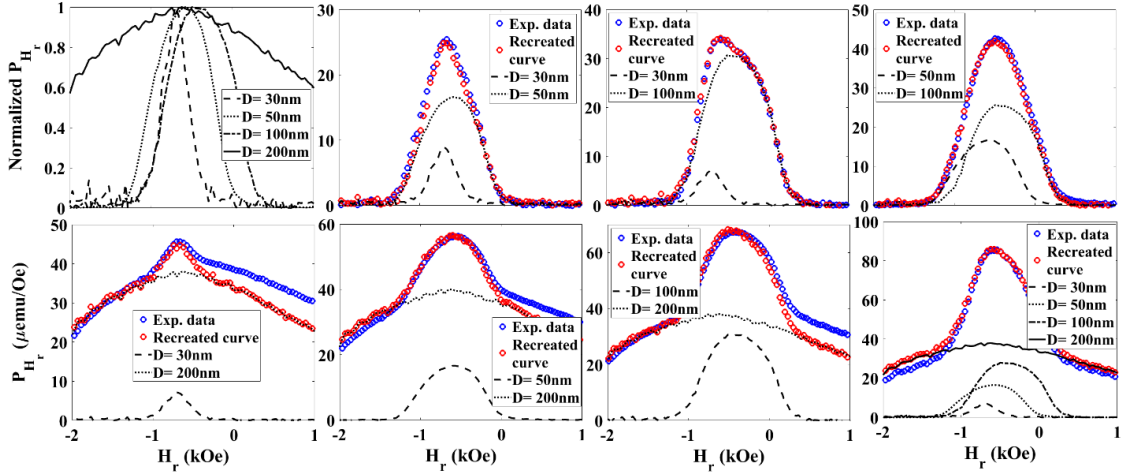

Figure 6: The projection of the FORC distributions on $H_r$-axis ($P_{Hr}$) results for individual and different combinations of the MNWs as indicated in the legend.

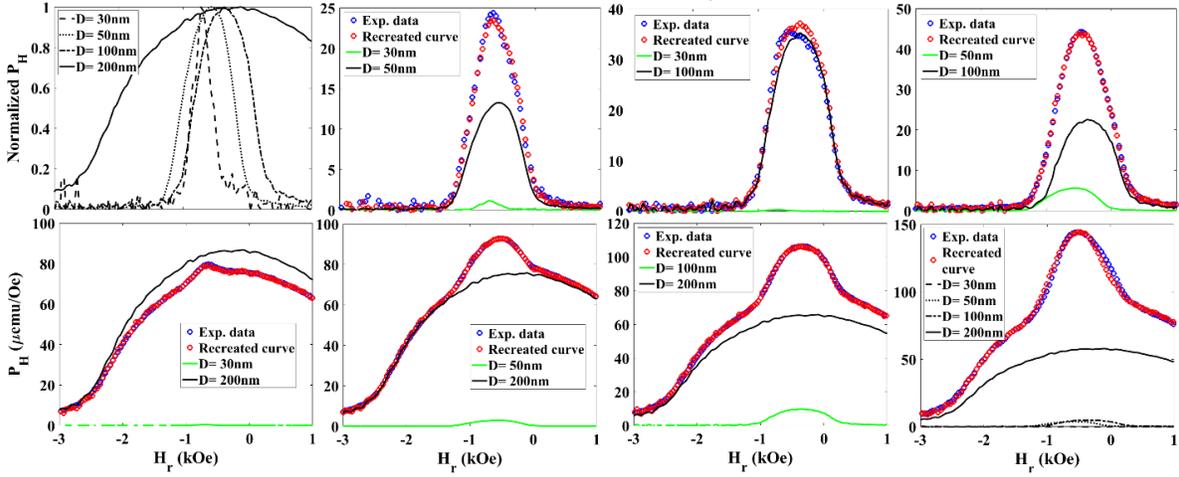

Figure 7: The projection of the FORC distributions on H-axis ($P_H$) results for individual and different combinations of the MNWs as indicated in the legend.

The $P_{Hr}$ maintains a sharp peak even for very small volume ratios, for example, the volume ratio of 1.3% for the 30nm and 200nm combination. That is because the $P_{Hr}$ not only depends on the MNWs coercivity and volume ratio, but also it depends on the irreversibility fraction (defined as irreversible magnetization to total magnetic moment). Note that the irreversibility fraction is a measure of I) how large an external energy (provided by external field, also known as Zeeman energy) is required to switch the magnetization direction, and II) how magnetically stable the magnetization direction will be once the switching occurs. For example, for non-interacting MNWs with very small diameters while magnetized along their easy axis, the reversal mechanism is coherent rotation, where the irreversibility fraction is 100%. On the other hand, for interacting MNWs with large diameters, the interaction field reduces the irreversibility fraction. Simply, this means the accuracy of the $P_{Hr}$ can be enhanced by controlling the irreversibility fraction of the MNWs in the combination, especially when the volume ratio is very small.

Figure 7 shows the BRM results of the individual type of the MNWs and their combinations. Since the samples have different magnetic moments, we normalized the BRM with respect to their saturation backfield remanence (remanence of the major hysteresis loop) to render them from -1 to +1. From Figure 5, it can be seen that the BRM of any combination is always valued between the BRMs of the individual MNWs in the combination. Therefore, the volume ratio can be determined by finding the BRM shift of the combination. Practically, two features characterize the BRM: I) the field where it is zero, which is average coercivity of the MNWs, and II) its slope, which is correlated to the MNW interaction field. Thus, the accuracy of the BRM can be enhanced by manipulating these two parameters.

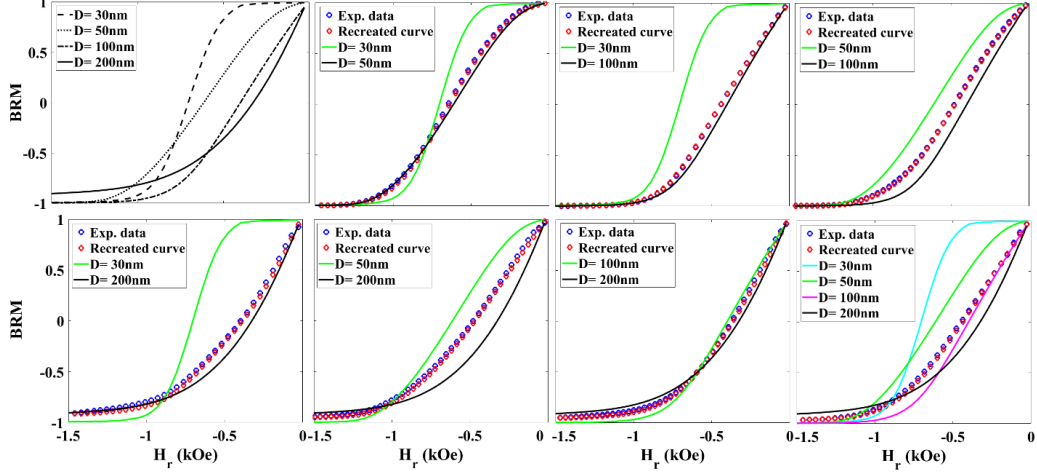

Figure 7: Backfield remanence magnetization (BRM) results for individual sample and different mixtures of MNW diameters as indicated in the legend.

Figure 8 provides the BRC distribution of the MNWs calculated by taking a derivative of the BRM with respect to the reversal field. As experimentally observed, the BRC of the individual type of the MNWs describes a maximum centered on the inflection points of the BRM curves as can be seen clearly in Figures 7 and 8. Practically, the behavior of the BRC of the combinations can be demonstrated by two parameters: I) the amplitude of each local peak (determined by the volume ratio of the MNWs), and II) the relative location of the peaks (indicating the coercivities of the MNWs). Therefore, similar to $P_{Hr}$, one can quantitatively describe the type and volume ratio of the MNWs in the combination by just analyzing these two features. It should be emphasized that the accuracy of the BRC depends on both of the amount and coercivity of component MNWs. Therefore, combinations of MNWs can be designed for optimal quantitative description by combining high coercivity and low coercivity types of the MNWs. For example, combination of 30nm diameter MNWs with 200nm diameter MNWs were easy to quantify with high accuracy due to completely distinct peaks in their BRC distributions.

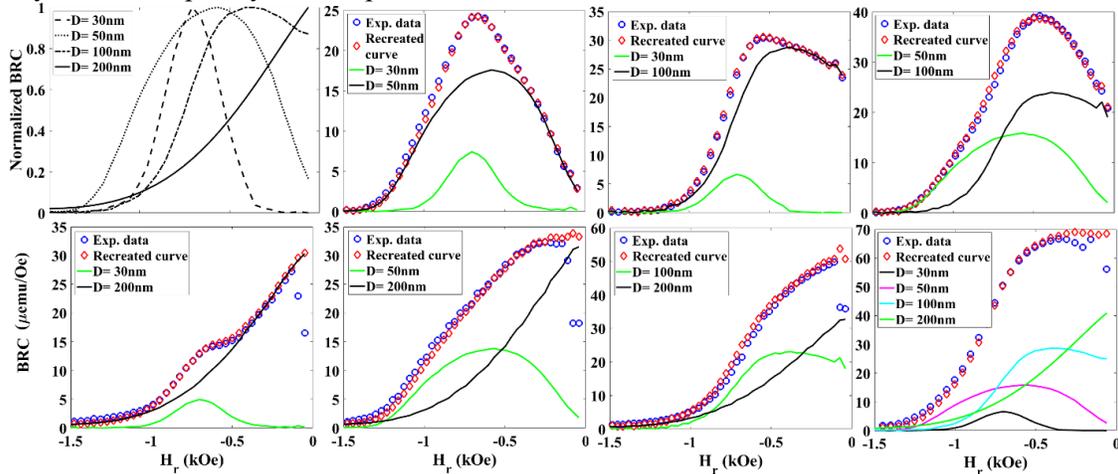

Figure 8: Backfield remanence coercivity (BRC) results for individual and different combination of the MNWs as indicated in the legend.

**Discussion**

Figure 9 summarizes the fitting results of each parameter, where Figure 9a shows the results for the combinations including two types of the MNWs and Figure 9b shows the results for the combination of all types. The fit volume ratio (χ) was found by minimizing the RMS error, and it is compared with the known χ. According to Figure 9, all parameters give the χ within a reasonable accuracy, the $P_{Hr}$ results seem to be slightly better compared to others. It could be expected that the $P_{Hr}$ and BRC have relatively similar results because they both depend on the coercivities, irreversibility fraction, and amount of each type of the MNWs. However, as it was already mentioned, there is a difference between the $P_{Hr}$ and BRC in which the $P_{Hr}$ shows the residual magnetization at the reversal curves while the BRC shows the residual magnetization at the zero applied field. Since the $P_{Hr}$ are measured at the reversal point, the applied field overcomes the interaction fields. Thus, the $P_{Hr}$ shows the residual magnetization purely dependent on the initial magnetization state without any influence of the interaction fields. On the other hand, since the applied field is zero while measuring the BRC, the interaction fields impact the magnetization, in which to minimize the magnetostatic energy, leading to random corrections on the residual magnetization. As a result, the BRC has a lower accuracy compared to the $P_{Hr}$. The accuracy of these parameters for the combination of all types, Figure 9b, highlights the advancement of this analysis in sensitive applications, such as enrichment and multiplexing of biological entities. For example, for targeting a specific cancer cell, it is essential to precisely estimate the amount of the cancer cell in order to determine the dosage of medicines.

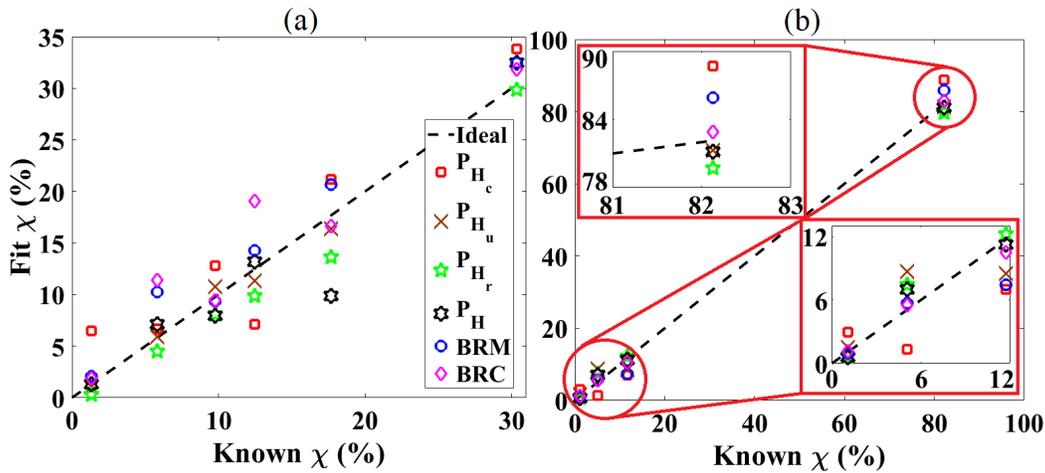

Figure 9: Comparing the quantitative results for each of the parameters. (a) The quantitative results for the combinations including two different types of the MNWs, and (b) the quantitative results for the combination including all four types of the MNWs. In subfigure (a), the $P_{Hu}$ predicts the volume ratio to be 52 for last one. For a better visualization of the results, we discard that point. In subfigure (b), the first, second, third, and fourth points show the volume ratio of 30nm, 50nm, 100nm, and 200nm MNWs to the total volume of the MNWs, respectively. The legend in both subfigures is the same.

**Conclusion**

In this work, we established a framework for quantitative data extraction and analysis using the FORC measurement. We showed that the $P_{Hc}$, $P_{Hu}$, $P_{Hr}$, $P_H$, BRM, and BRC can be readily extracted for quantitative analysis of the MNWs, something that cannot be preserved from the conventional representation of the FORC data. Our experimental observation indicates that the $P_{Hr}$ has slightly a better capability for quantifying the volume ratio (χ) of the MNWs because it employs the effects of the irreversibility while discarding external effects, such as interaction fields. Furthermore, these parameters are able to estimate the χ of the individual types of the MNWs in a combination containing several types of the MNWs within a reasonable accuracy. This finding opens numerous opportunities in biomedical applications by ceasing the quantification of several biological entities, such as cancer cells, for achieving an effective medical treatment.